%% file: main.tex
\documentclass[sigconf]{acmart}
\AtBeginDocument{%
  }

\acmConference[CHI '26 BiAlign Workshop]{Make sure to enter the correct
  conference title from your rights confirmation email}{April 15,
  2026}{Barcelona, Spain}

\begin{document}

\title{Rationalize: Shared Semantic Reasoning for Human-AI Alignment}


\author{Aritra Dasgupta, Naga Datha Saikiran Battula, Avina Nakarmi, \\ Sohom Sen, Subhodeep Ghosh \& Xun Song}
\affiliation{%
  \institution{New Jersey Institute of Technology}
  \city{Newark}
  \country{USA}}






\renewcommand{\shortauthors}{Dasgupta et al.}

\begin{abstract}
We introduce Rationalize, a role-pair framework for shared semantic reasoning between humans and AI models in data-driven sensemaking. Building on ideas in human-machine teaming and critical thinking, we conceptualize human-AI interaction as a series of complementary role pairs (Explorer-Guide, Investigator-Informant, Teacher-Student, Judge-Advocate) operating in a shared reasoning space. In this space, human analysts and AI models (such as LLMs) make purposes, questions, assumptions, evidence, inferences, and implications explicit, facilitating alignment not only at the output level but at the level of rationalization of intent and action by each side. We relate these role pairs to the bidirectional human-AI alignment framework, illustrating how “aligning AI to humans” and “aligning humans to AI” differ by role, and sketch a collaborative research agenda for alignment design and assessment using element-level and role-specific approaches.
\end{abstract}



\maketitle

\section{Introduction}

\input{intro.tex}

\input{bidirectional}

\input{mapping}
\bibliographystyle{ACM-Reference-Format}
\bibliography{CHI26}

\end{document}

%% file: intro.tex
Successful human-AI collaboration depends on two reasoning agents, human and machine, working together to reach a shared understanding, even though they cannot see each other's internal thought processes. Instead of focusing on better interfaces or models, we suggest creating a shared semantic reasoning space. This is a clear and structured way for both sides to express, discuss, and negotiate key parts of their reasoning, like purpose, assumptions, evidence, inferences, and implications, as equals.

AI-powered data analytics and large language models (LLMs) are changing how people tackle data-driven sensemaking. However, most systems still use AI in a one-way process: humans ask questions, and AI gives answers. As a result, sensemaking happens around the AI, not with it~\cite{shen2024towards}.

We suggest a role-based approach to human-AI collaboration in sensemaking, built around shared reasoning. This approach uses the roles of Explorer, Investigator, Teacher, and Judge~\cite{wenskovitch2021beyond}, and adapts them for LLM-in-the-loop situations by pairing each human role with an AI counterpart (Guide, Informant, Student/Tutor, Advocate). Each pairing forms a unique role configuration, which is a specific way to enter the shared reasoning space, each with its own responsibilities and focus. To organize this space, we use Paul and Elder’s critical thinking framework~\cite{paul2019miniature,louisvilleElements}, which defines eight Elements of Thought that both humans and AI can share and discuss.

We argue that when both sides clearly lay out these elements, human-AI interaction becomes true collaboration instead of just a transaction. This is achieved through role pairs. Each role pair focuses on certain elements: for example, purpose and perspective in exploration, evidence and inference in investigation, concepts and assumptions in teaching, and implications and accountability in judging.

Our main contribution is a conceptual and design-focused framework that (1) defines a shared semantic reasoning space based on critical thinking, (2) links human-AI role pairs to the key elements in each phase of sensemaking, and (3) explains how two-way alignment: adjusting AI to human goals and helping people understand AI reasoning, changes across different role configurations.

%% file: bidirectional.tex
\section{Bidirectional Alignment for Human-AI Data Sensemaking}

Traditional analytics workflows have long relied on a familiar split: machines take care of large-scale computation and retrieval, while humans focus on forming hypotheses and making sense of results~\cite{thomas2006visual}. Visual analytics built on this idea through interactive systems designed to help users ``detect the expected, discover the unexpected,'' while leaving interpretation, uncertainty assessment, and trust decisions in human hands. However, as Wenskovitch et al.\ point out, this arrangement is becoming less tenable. With advances in machine learning and conversational AI, systems are no longer passive tools; they can generate hypotheses, suggest questions, and learn from interaction, increasingly acting as collaborators rather than instruments~\cite{wenskovitch2021beyond}.

To better describe this shift, Wenskovitch et al.\ outline four roles that humans take on in human–machine collaboration: \emph{Explorer} (open-ended search and idea generation), \emph{Investigator} (targeted analysis and verification), \emph{Teacher} (providing feedback to improve models), and \emph{Judge} (evaluating outcomes and decisions)~\cite{wenskovitch2021beyond}. Visual analytics systems support these roles through mixed-initiative interactions such as recommending data, suggesting labels, explaining results, or soliciting feedback. At the same time, this work highlights a growing tension around agency and shared understanding: as AI takes on more of the search and reasoning work, people may feel less in control, making it harder to monitor errors or decide when to trust the system~\cite{wickens2015complacency}.

Recent work on human–AI alignment offers a broader lens for understanding how people and AI systems adapt to one another over time. Shen et al.\ describe \emph{bidirectional alignment} as the interplay between two processes: aligning AI systems with human goals and values, and helping humans understand and work effectively with AI systems~\cite{shen2024towards}. They highlight three persistent challenges: alignment must be treated as an ongoing, evolving process rather than a one-time goal; human values and intentions are complex and difficult to model; and interaction design must support mutual adaptation rather than simple control.

From our perspective, the key takeaway is that bidirectional alignment is ultimately about how reasoning is shared, surfaced, and revised—not just about controlling outputs. Data-driven sensemaking~\cite{klein2007data,pirolli2005sensemaking}, with its cycles of exploration, hypothesis formation, explanation, and decision-making under uncertainty, provides a natural setting to study these dynamics. The role pairs introduced in the next section can be understood as structured ways of making this shared reasoning visible and usable at different stages of the sensemaking process.

\subsection{Characterizing Shared Semantic Reasoning}

The critical thinking framework~\cite{paul2019miniature} provides a concise and practical way to describe reasoning through eight Elements of Thought: (1) purpose, (2) question at issue, (3) information (data, evidence), (4) concepts and ideas, (5) assumptions, (6) inferences and interpretations, (7) implications and consequences, and (8) point of view~\cite{louisvilleElements}.

These elements serve as the basic vocabulary of a shared reasoning space. When both human and AI participants can make these elements explicit—stating their goals, revealing assumptions, and laying out inferential steps—it becomes easier to see where their reasoning differs, which differences matter, and how they might be resolved. Without this shared structure, alignment remains implicit and fragile, with each side reacting to outputs without insight into how they were produced.

From a sensemaking standpoint, these elements apply to both human and AI reasoning. A human analyst might define a goal (for example, understanding a spike in defect rates), pose a question, gather evidence, and use domain knowledge and assumptions to arrive at a conclusion with certain implications from a particular perspective~\cite{louisvilleElements}. An AI system, in parallel, operates with its own counterparts: objectives and constraints (purpose), prompts or tasks (question), training and retrieved data (information), learned representations (concepts), inductive biases (assumptions), inference procedures (conclusions), and implicit trade-offs (implications and point of view).

This parallel structure matters. It turns the Elements of Thought into a shared language for interaction rather than a checklist for human reasoning alone. Through this lens, aligning AI to humans involves giving users ways to inspect and adjust system goals, assumptions, and reasoning patterns. Aligning humans to AI involves helping users understand how the system’s data, concepts, and perspective differ from their own, and when those differences should prompt trust, skepticism, or intervention.

\section{Role-Based Human–AI Pairings}

Role pairs provide a way to enter and work within this shared reasoning space. Rather than treating roles as fixed identities, we view them as configurations that highlight different aspects of reasoning. Each pairing brings certain Elements of Thought to the foreground, introduces specific risks of misalignment, and calls for different forms of interaction. These configurations are not linear stages; teams often move between them repeatedly within a single session.

Building on the roles identified by Wenskovitch et al.—Explorer, Investigator, Teacher, and Judge—we extend them to settings where LLMs are active participants~\cite{wenskovitch2021beyond}. This results in four pairings: Explorer–Guide, Investigator–Informant, Teacher–Student, and Judge–Advocate. Each pairing emphasizes different elements and different forms of alignment. Making these distinctions explicit helps designers anticipate where breakdowns are likely and how to support more effective collaboration.

\subsection{Explorer–Guide}

In the Explorer–Guide pairing, the human leads open-ended exploration under uncertainty, while the AI suggests directions to pursue. This is often a ``cold-start'' phase, where the problem is not yet well defined and broad exploration is valuable~\cite{thomas2006visual}. The human gradually shapes a sense of purpose and formulates tentative questions, often starting from incomplete understanding.

The AI Guide supports this process by proposing datasets, views, patterns, and hypotheses. Rather than simply executing queries, it acts as a partner in suggesting where to look next. Here, the shared reasoning space centers on purpose, question, and point of view. Both sides must converge on what counts as interesting or worth exploring, and the AI must make its internal criteria—such as similarity measures or ranking priorities—understandable to the user.

Alignment in this setting works in both directions. On one hand, the system adapts to the user’s evolving preferences, such as whether they prioritize anomalies or trends. On the other hand, users develop a clearer sense of what the system is likely to surface and where its biases lie. Effective interfaces make these assumptions visible and clarify trade-offs such as novelty versus plausibility or coverage versus precision.

Systems like Selenite~\cite{selenite} illustrate this dynamic by combining global overviews with local annotations, allowing users to reshape the structure suggested by the AI. Similarly, SenseMate~\cite{10.1145/3640543.3645194} and LLM4Vis~\cite{wang-etal-2023-llm4vis} guide users through unfamiliar tasks with explanations that support understanding rather than blind acceptance. In all these cases, the AI contributes to a shared process of inquiry rather than delivering final answers.

\subsection{Investigator–Informant}

As analysis becomes more focused, the interaction shifts to the Investigator–Informant pairing. The human asks targeted questions and evaluates answers, while the AI gathers evidence, performs analysis, and communicates results along with their uncertainty.

The shared reasoning space here revolves around information, inference, and implications. Crucially, the AI must do more than provide answers—it must make its reasoning visible by showing what data it used, how it drew conclusions, and what alternative interpretations might exist. This transparency allows the human to question, refine, or extend the analysis.

Alignment requires the system to present explanations in ways that match the user’s standards for evidence and reasoning. At the same time, users need support in understanding when the system’s conclusions are reliable and when they may be limited by data or modeling choices.

Modern systems such as Snowy~\cite{snowy}, Boomerang~\cite{boomerang}, Talk2Data~\cite{Talk2Data} move beyond simple question answering by suggesting relevant queries and exposing intermediate reasoning steps. 

Tools like InReAcTable~\cite{inreactable} and Visual (dis)Confirmation~\cite{visual_disconfirmation} further support this process by making analytical pathways and discrepancies visible. Across these systems, the most effective designs preserve the chain of reasoning rather than compressing it into a single answer.

\subsection{Teacher–Student}

The Teacher–Student pairing captures situations where humans shape AI behavior through feedback, or where AI helps humans learn. In the Teacher role, the human provides labels, corrections, and explanations to guide the system’s learning. In the Student–Tutor configuration, the roles are reversed, with the AI helping the human refine their understanding.

Here, the shared reasoning space focuses on concepts, assumptions, and information. Teaching is not just about outcomes; it is about reshaping how the system represents and interprets the world. For this to work, the system must make its internal changes visible, allowing the human to see how their feedback has influenced its behavior.

Alignment involves both incorporating structured human feedback into the model and helping users understand how their input affects the system. Without this transparency, users may feel that their efforts have little impact, leading to frustration and reduced trust. Interfaces can address this by showing how decision boundaries shift or summarizing what the system has learned.

In Student–Tutor settings, the emphasis shifts to supporting human reasoning. The AI can prompt users to clarify goals, question assumptions, and consider alternative perspectives~\cite{paul2019miniature,louisvilleElements}. This makes human reasoning more visible and open to revision, turning the interaction into a joint learning process.

\subsection{Judge–Advocate}

In the Judge–Advocate pairing, the focus is on evaluation and accountability. The human Judge assesses the results, particularly in high-stakes contexts, while the AI Advocate presents evidence, trade-offs, and alternative options.

The shared reasoning space here centers on implications, assumptions, and points of view. Disagreements are often not about facts but about values—whose interests matter, which risks are acceptable, and what trade-offs are justified. The system must therefore make these value-laden aspects explicit so that the human can deliberate rather than defer.

Alignment involves both encoding appropriate value priorities in the system and helping users understand its capabilities and limits. The role of AI is to present structured reasoning that informs judgment, not to make final decisions. The human, in turn, must treat AI output as one input among many in a broader decision process.

\medskip
Taken together, these role pairs show that different stages of sensemaking emphasize different elements of reasoning: purpose and perspective in exploration, evidence and inference in investigation, concepts and assumptions in teaching, and implications and accountability in evaluation. By making these differences explicit, the role-based framework provides a way to design interactions that support alignment where it matters most.

\begin{table*}[t]
  \centering
  \small
  \setlength{\tabcolsep}{3pt}
  \begin{tabular}{p{2.6cm} p{1.8cm} p{2.2cm} p{2.6cm} p{2.4cm} p{2.6cm} p{2.6cm}}
    \toprule
    \textbf{Role Pair} &
    \textbf{Purpose} &
    \textbf{Question at Issue} &
    \textbf{Information (Data, Evidence)} &
    \textbf{Assumptions} &
    \textbf{Inferences \& Conclusions} &
    \textbf{Implications \& Point of View} \\
    \midrule

    Explorer--Guide
    &
    Co-articulate exploration goals (e.g., novelty, risk, coverage); AI operationalizes ``interestingness'' via recommendation objectives.
    &
    Evolve from vague to sharper questions; AI helps surface candidate questions and alternative framings.
    &
    Guide proposes candidate datasets, features, and views; Explorer vets relevance and coverage of suggested information.
    &
    Make default priors and search biases explicit; Explorer corrects or constrains them.
    &
    Early, tentative inferences; AI can propose speculative hypotheses with appropriate hedging.
    &
    Negotiate perspectives (e.g., user, stakeholder, organization); clarify what counts as valuable or problematic outcomes. \\[0.4em]

    Investigator--Informant
    &
    Focus on resolving specific ``known unknowns''; clarify analytic sub-goals and decision criteria.
    &
    Precisely formulate investigative questions; AI supports reformulation and decomposition.
    &
    Informant gathers, aggregates, and ranks evidence; human curates sources, quality, and sufficiency.
    &
    Surface model and analyst assumptions (e.g., independence, stationarity, causal structure) for scrutiny.
    &
    Jointly assess whether conclusions follow from evidence and domain logic; inspect failure modes.
    &
    Examine downstream consequences of acting on conclusions; foreground affected stakeholders and risks. \\[0.4em]

    Teacher--Learner / Student--Tutor
    &
    Teacher's purpose is to shape model behavior; Student's purpose is to gain understanding and skill.
    &
    Teacher frames teaching tasks (what the model should learn); Tutor elicits learner questions and gaps.
    &
    Curate training/feedback data; highlight where more examples or counterexamples are needed.
    &
    Identify and correct problematic priors or heuristics on both sides (e.g., shortcuts, stereotypes).
    &
    Analyze how both model and human derive conclusions; critique reasoning chains and rationales.
    &
    Discuss the implications of learned behavior (e.g., safety, fairness); surface value tensions and normative commitments. \\[0.4em]

    Judge--Advocate
    &
    Judge's purpose is oversight and accountability; Advocate's purpose is to support evaluation with structured evidence.
    &
    Formulate evaluative questions (e.g., ``Is this deployment acceptable?''); consider multiple review lenses.
    &
    Advocate assembles evidence about performance, failures, and context; Judge probes for missing or biased information.
    &
    Make explicit institutional, legal, and ethical assumptions embedded in metrics and policies.
    &
    Evaluate whether actions and outcomes are justified; compare alternative decision rules or models.
    &
    Examine short- and long-term consequences across stakeholders; contrast points of view and value trade-offs. \\
    \bottomrule
  \end{tabular}
  \caption{Conceptual role--element matrix linking human--AI role pairs to selected Elements of Thought~\cite{paul2019miniature,louisvilleElements}. Each cell indicates where bidirectional alignment work is concentrated for that role configuration.}
  \label{tab:role_element_matrix}
  \vspace{-.2in}
\end{table*}

%% file: mapping.tex
\section{The Shared Reasoning Space in Practice}

The role--element matrix (Table~\ref{tab:role_element_matrix}) is more than an analytic device; it can also guide the design of the shared reasoning space. Once a role pair is active, the matrix helps identify which reasoning elements should be surfaced in the interface, what kinds of feedback matter most, and where breakdowns are likely to occur. We discuss three design implications below.

\subsection{Designing for Explicit Expression of Elements}

A shared reasoning space only works if both participants can actively contribute to it. One of the clearest implications of the role--element matrix is that interfaces should help people and AI systems explicitly express the reasoning elements that matter in a given interaction, rather than leaving them buried in open-ended conversation.

For an Explorer--Guide interaction, the interface might prompt the human to specify purpose, questions, and constraints, while the AI is expected to make its assumptions and alternative perspectives visible. For an Investigator--Informant interaction, response formats might be designed to consistently reveal the information used, the inferences made, and the implications that follow.

Consider the Explorer--Guide case. The Explorer’s purpose may be fluid, layered, and hard to pin down: they may be seeking novelty, insight, and risk reduction at the same time. By contrast, the Guide may operationalize purpose more narrowly, for example by prioritizing predicted relevance or information gain. The interface should help connect these two views. It should allow the Explorer to express their goals in natural language while requiring the Guide to show how it interpreted those goals and how that interpretation shaped its suggestions.

A similar need arises in the Investigator--Informant pairing. Misalignment may stem from the sources the Informant relies on, the reasoning shortcuts it takes, or the consequences it fails to address. The interface should therefore give the Investigator ways to respond at the level of reasoning itself: to mark a source as unreliable, to question an inference strategy, or to request that certain consequences be included in the answer, such as worst-case outcomes or impacts on underrepresented groups.

\subsection{Element-Referenced Feedback}

Instead of reducing feedback to thumbs up or thumbs down, systems built around a shared reasoning space could support \emph{element-referenced feedback}. In this model, users do not simply reject an output; they identify what part of the reasoning is problematic, for example by saying that an assumption is incorrect, the information is incomplete, an implication is missing, or the point of view does not match their own.

This matters for two reasons. First, it produces more informative feedback. Rather than merely signaling failure, it specifies the nature of the failure. Second, it gives users a way to express their own standards of reasoning, not just their approval or disapproval of the result. Over time, records of this kind of feedback could become a valuable source of evidence about how bidirectional alignment develops in different role configurations, and could support empirical study of alignment in real sensemaking settings~\cite{shen2024towards}.

This is especially important in Teacher--Student settings. If the AI relies on concepts the human does not endorse, or if the human is working from assumptions the AI cannot properly represent, the teaching process starts to break down. In such cases, the Teacher or Tutor should be able to respond with feedback such as ``your concept of X is too broad'' or ``you are assuming that Y is independent of Z,'' rather than simply marking the answer as wrong. Without this level of specificity, the substance of teaching disappears inside the model, and the human has no reliable way to tell whether the intended conceptual change has actually taken place.

\subsection{Metacognitive Scaffolding and Evaluation}

The shared reasoning space is not fixed. It changes over time as the human and the AI revise their goals, interpretations, and judgments. Interfaces can support this process by making those shifts visible. For example, they might show how an Explorer’s goals have evolved, how the Informant’s evidence base has changed, or how a Judge--Advocate pair has refined its understanding of consequences during deliberation. Such views can support metacognition by helping people reflect on their own reasoning and notice when AI influence has been useful, misleading, or excessive.

From the standpoint of human adaptation, this kind of scaffolding speaks directly to one of the central concerns raised by Shen et al.: how to help people understand, critique, and adapt to AI systems over time. The shared reasoning space provides the ground on which that kind of reflection becomes possible.

The role--element matrix also suggests a path toward more systematic evaluation of bidirectional alignment. For each role pair, one could imagine metrics that track alignment in how purposes are framed, how often the same sources are used, how frequently inferences conflict, or how differently consequences are assessed. These measures could help distinguish productive disagreement from genuine alignment failure. In the former case, the disagreement is visible and available for discussion within the shared reasoning space. In the latter, the breakdown remains hidden. That distinction only becomes possible if the system is designed to expose reasoning in the first place, which is one of the central commitments of the Rationalize framework.

\section{Conclusion: Toward a Collaborative Research Agenda}

We began from a simple premise: meaningful alignment between humans and AI requires visibility into reasoning. The Rationalize framework is an attempt to make that possible. By grounding collaboration in a shared semantic reasoning space, and by organizing that space through elements of thought and role-based pairings, the framework makes the reasoning of both human and AI participants more visible, more comparable, and more open to correction. The role pairs are not just an organizing device. They are what make the shared space concrete. Each pair brings different reasoning elements and different alignment risks to the forefront: purpose and perspective in exploration, evidence and inference in investigation, concepts and assumptions in teaching, and implications and accountability in judgment.

This is not presented as a finished solution. Quite the opposite. It is a conceptual framework that will only become meaningful through sustained work across research communities.

\textbf{For the AI and machine learning community:}
A central technical challenge is to build models that can expose their reasoning in ways that are genuinely useful to human collaborators. The goal is not post-hoc rationalization, but clearer access to what the system is trying to accomplish, what information it is relying on, what assumptions it is making, and what consequences it is implicitly following. Although current work on interpretability and chain-of-thought reasoning moves in related directions, it is often aimed more at researchers inspecting models than at people collaborating with them. A shared reasoning space calls for models that can respond to targeted feedback such as ``this assumption is wrong'' or ``this inference violates domain logic,'' and revise the relevant part of the reasoning process rather than treating all feedback as a global correction signal. This points toward new architectures for structured feedback, new training objectives that reward transparency, and new evaluation approaches that assess the quality of the reasoning made available in the shared space, not only the final answer.

\textbf{For the HCI and data analytics community:}
The design challenge is to bring the shared reasoning space to the foreground of interaction. That means moving beyond interfaces that focus only on outputs and toward interfaces that let people inspect and shape the reasoning behind those outputs. Systems that recognize role pairs and adapt prompts, explanations, and feedback mechanisms depending on whether the interaction is exploratory, investigative, instructional, or evaluative would be an important step forward. So would interfaces that preserve a history of reasoning from both sides, allowing users to examine how the shared space changes over time. By studying which elements people attend to, which they ignore, and how these patterns shift with expertise, researchers can ground the framework in observed behavior rather than abstract design principles alone.

\textbf{For the responsible AI and policy community:}
The Judge--Advocate pairing highlights a dimension of alignment that technical methods alone cannot resolve: whose values, interests, and perspectives are represented in the reasoning space. A system may be interactive and transparent, yet still fail to be genuinely shared if it reflects only a narrow range of viewpoints. Addressing this requires participatory approaches in which affected communities help shape the reasoning space itself, not merely evaluate system outputs after the fact. They should have a voice in determining which elements matter, which consequences deserve attention, and which points of view need representation.

\textbf{A shared agenda:}
What ties these directions together is a common emphasis on reasoning, rather than behavior, output, or preference alone, as the basis for alignment. Much current alignment work is behavioral: it observes what models do and attempts to steer them toward preferred outcomes. A shared reasoning space suggests a more structural approach. It asks what kind of epistemic relationship between people and AI could support durable, meaningful alignment, and then designs for that relationship rather than for a single result.

The immediate design problems are substantial, and so are the long-term stakes. A form of collaboration in which humans and AI can inspect, challenge, and learn from one another’s reasoning is not just better aligned. It points to a more dependable and more substantive model of collaboration altogether.

\section{Acknowledgement}
This work was funded in part by the National Science Foundation-
tion (NSF) grants 2312932 and 2326195.